\documentstyle[aps,prl,preprint]{revtex}
\def\fun#1#2{\lower3.6pt\vbox{\baselineskip0pt\lineskip.9pt
        \ialign{$\mathsurround=0pt#1\hfill##\hfil$\crcr#2\crcr\sim\crcr}}}

\begin{document}
%\vspace*{-62pt}

\vspace{0.5in}

{\title{\vskip-2truecm{\hfill {{\small  TURKU-FL-P19-95\\
        \hfill SISSA/AP/95/112\\
        \hfill hep-ph/9509252\\
        }}\vskip 1truecm} {\bf ON THE NONEQUILIBRIUM EFFECTIVE POTENTIAL}}

\vspace{1.cm}

\author{ANTONIO RIOTTO\thanks{riotto@tsmi19.sissa.it. Address after
November 95: Theoretical Astrophysics Group, NASA/Fermilab, Batavia,
IL60510, USA.}$^{1)}$ and IIRO VILJA\thanks{vilja@utu.fi.}$^{2)}$}

\vspace{1.0cm}

\address{ $^{1)}$ International School for Advanced Studies, SISSA-ISAS,\\
Strada Costiera 11, I-34014, Miramare, Trieste, Italy\\and\\
Istituto Nazionale di Fisica Nucleare, Sezione di Padova, 35100 Padova,
Italy}

\address{ $^{2)}$Department of Physics, University of Turku, FIN-20500
Turku, Finland}

\maketitle

\vspace{1.cm}

\begin{abstract}
\baselineskip 16pt
Nonequilibrium phenomena of the phase transitions are studied. It
is shown that due to finite relaxation time of the particle distributions,
the use of scalar background dependent distribution functions is inconsistent.
This observation may change the picture of rapid processes during the
electroweak phase
transition, like subcritical bubble formation and propagation of bubble
walls.

\end{abstract}

\thispagestyle{empty}

\newpage
\pagestyle{plain}
\setcounter{page}{1}
\def\beq{\begin{equation}}
\def\eeq{\end{equation}}
\def\beqa{\begin{eqnarray}}
\def\eeqa{\end{eqnarray}}
\def\tr{{\rm tr}}
\baselineskip 20pt

The possibility that the Universe went through a series of phase
transitions during its expansion and cooling down from temperatures
close to the Planck scale has been intensively investigated in the past
\cite{kt}. It is a common hope that many of the current questions
of cosmology can be answered by studying the
nontrivial dynamics of the approach to equilibrium in complex systems.
Nevertheless, despite their immense relevance, only very recently
more effort has been made
to understand nonequilibrium phenomena occured in the early Universe.

Thermalization, reheating and relaxation are nonequilibrium phenomena that
play crucial role
in the completion of the inflationary era, which is  thought to
solve the horizon and the homogeneity problems \cite{inflation} of modern
cosmology. Recent
investigations on the non-linear quantum dynamics of scalar fields
have implications for the reheating and reveal that particle
production appears to be significantly different from linear estimates
due to the time evolution of the inflation field \cite{noneq}. The
quantum non-linear effects lead to an extremely effective dissipational
dynamics and particle production even in the simplest self-interacting
theory where the single particle decay is kinematically forbidden. Also they
might help in alleviating the so-called Polony \cite{polony}
problem concerning flat
directions for some of the moduli fields in string theories.

At the electroweak scale, the focus has been in generation of the baryon
asymmetry during a first-order phase transition where
the $SU(2)_L\otimes U(1)_Y$ symmetry  is broken down to
$U(1)_{em}$ \cite{ew}. It is currently
believed that nonequilibrium conditions are a crucial ingredient for
electroweak baryogenesis, even though there are certain questions
related to the reliability of the perturbative expansion for weak
transitions \cite{quiros}. Moreover, the  mechanism of the weak
first-order transitions \cite{sb} is still an open question.
Consider
models with double-well potentials where the system starts localized
on one minimum. For sufficiently weak transitions subcritical
bubbles of the other phase could be thermally nucleated, giving rise to
{\it e.g.} an effective phase mixing between the two available phases before
cooling down to the tunneling temperature at which critical
bubbles are expected to be nucleated. This may have dramatic
consequences for any electroweak baryogenesis mechanism.

So far, however, most done to study the nonequilibrium aspects of phase
transitions, {\it e.g.} the influence of  subcritical bubbles on a
first-order electroweak phase transition, is strictly related to the
equilibrium finite temperature field theory \cite{ka}
and effective potential  which, by its very definition, is only adequate
to describe equilibrium situations.
%Calculations are usually done
Two equivalent methods are usually applied: the imaginary-time formalism
(in Euclidean time)  or the real-time
formalism with {\it equilibrium} distribution functions.
They are used to obtain the thermal part of the equilibrium effective
potential by integrating out all the fermionic and bosonic
degrees of freedom of the theory with $\phi_c$-dependent masses.

In the equilibrium case the finite temperature part of the one-loop
effective potential reads \cite{dolan}
(besides an additive $\phi_c$-independent constant)
\begin{equation}
\label{pot}
V_{{\rm eq}}(T,\phi_c)=\sum_i\:\int\:\frac{d^3 k}{(2\pi)^3}\:
\int_{\infty}^{\omega_{{\bf k}}(m_i)}\:d\omega\:f^{i}_{{\rm
eq}}(\omega),
\end{equation}
where $\omega_{{\bf k}}(m_i)=\sqrt{|{\bf k}|^2+m^2_{i}(\phi_c)}$,
$m_i(\phi_c)$ is the eigenstate of the mass matrix corresponding to the
$i$-th degrees of freedom in a $\phi_c$-dependent background.
%and $g_i$'s properly count all the degrees of freedom in the thermal bath.
Note, that $f^i_{{\rm eq}}$
is the usual equilibrium Fermi-Dirac or Bose-Einstein distribution
function. The effective potential $V_{{\rm
eq}}(T,\phi_c)$
represents the {\it equilibrium} free energy density as a function of
the classical order parameter $\phi_c$ and  can be used to
determine the nature of the phase transition and static quantities like
the critical temperature, but it is not the appropriate tool for the
description of real-time processes, which are crucial to understand the
mechanism by which the system approaches the equilibrium \cite{mi}.

The aim of the present Letter is to point out that
the use of the equilibrium finite temperature effective potential
becomes completely
unreliable when the thermalization time $\tau_{{\rm th}}$ of
the degrees of freedom getting a $\phi_c$-dependent mass in the thermal bath
is larger than the typical time scale
$\left|\phi_c/\dot{\phi_c}\right|$, {\it I.e.} when
the macroscopic order parameter $\phi_c$ does not vary slowly in
comparison with $\tau_{{\rm th}}$.  More generally, we can state that
the equilibrium
effective potential is not trustable at times $t$ smaller than the
thermalization time $\tau_{{\rm th}}$
in a $\phi_c$-background changing
in space and time at a generic instant $t=0$.

 To avoid any confusion, we
observe here that what we call thermalization time  should be understood
as the time needed for the degrees of freedom with a $\phi_c$-dependent
mass to feel the change of the background $\phi_c$ {\em and} to relax their
distribution functions to their equilibrium values $f^i_{{\rm eq}}$ with
the appropriate mass $m_i(\phi_c)$ \cite{weldon}. Thus, for instance, the
thermalization time for fermions is dictated by the Yukawa
interactions with the scalar field $\phi$ and not by the strong
interactions with gluons in the thermal bath, which are much faster.

In this paper we propose a nonequilibrium approach to describe the
properties of the system in the specific limits depicted above and
apply it to the  Standard Model (SM) effective
potential. We
will also briefly discuss some physical situations in which our simple, but
crucial,
statement might have interesting consequences.

Let us first briefly clarify  with a simple example our observation.
Suppose a universe filled only with a real scalar field $\phi$
and a fermion field $\psi$ whose Lagrangian density reads
\begin{equation}
{\cal L}=\frac{1}{2}\left(\partial_\mu
\phi\right)^2+i\bar{\psi}\not\!\partial\psi
-V(\phi)-g\phi\bar{\psi}\psi.
\end{equation}
The zero temperature tree level potential $V(\phi)$ is given by
\begin{equation}
V(\phi)=-\frac{1}{2}\mu^2\phi^2+\frac{1}{4}\lambda \phi^4.
\end{equation}
At high temperatures the effective
potential for the classical field $\langle\phi\rangle=\phi_c$
has the form
\begin{equation}
V(T,\phi_c)=\frac{1}{2}m^2(T)\phi_c^2+\frac{1}{4}\lambda(T)\phi_c^4,
\end{equation}
where
\begin{equation}
m^2(T)=-\mu^2+\left(\frac{\lambda}{4}+\frac{g^2}{6}\right)T^2,
\end{equation}
is the plasma mass for the $\phi_c$-field whereas $\lambda(T)$ receives only
logarithmic one-loop corrections.
Above the critical temperature $T_c$, the free
energy of the system is minimized by $\phi_c=0$ and the
Fermi-Dirac and Bose-Einstein distribution functions for the fermionic
and bosonic  degrees of freedom of the
heat bath are the equilibrium ones
\begin{eqnarray}
f^{{\rm eq}}_\psi({\bf k},\phi_c=0)&=&\left(1+{\rm e}^
{\beta\left|{\bf k}\right|} \right)^{-1}\nonumber\\
f^{{\rm eq}}_\phi({\bf k},\phi_c=0)&=&\left({\rm e}^{\beta \sqrt{\left|
{\bf k}\right|^2 + m^2(T)}}-1\right)^{-1}.
\end{eqnarray}

Let us now imagine that  at a generic  time $t=0$ in a certain region of space
the vacuum expectation value  becomes different from zero,
$\phi_c=\phi_c({ \bf x},t)$. This
can happen because in that region the system suffers a thermal
fluctuation or simply because the temperature of the thermal bath has cooled
down to the critical temperature $T_c$, $\phi_c=0$ becomes an unstable
point and $\phi_c$ starts to roll down.
The equation of motion for $\phi_c({\rm \bf x},t)$ reads
\begin{equation}
\Box\phi_c({\bf x},t)+\frac{\partial V\left[\phi_c,f_\psi,
f_\phi\right]}{\partial\phi_c}=0,
\end{equation}
where $V(\phi_c)$ depends upon the distribution functions $f_\psi$ and
$f_\phi$ through the one-loop corrections. The question is now whether
we can use, for instance,  the {\it equilibrium} distribution function
$f^{{\rm eq}}_\psi(\phi_c)$  with a
$\phi_c$-dependent mass  $m_\psi(\phi_c)=g\phi_c$
to calculate the contribution of the fermionic degrees of freedom
to the one-loop effective potential.

To answer the question one should solve the the Boltzmann equation
\begin{equation}\label{bol}
\partial_t f_\psi + \dot{{\bf x}}\cdot\nabla f_\psi+\dot{{\bf k}}\cdot
\partial_{{\bf k}} f_\psi= {\cal C}[f_\psi ],\nonumber
%-\frac{1}{\tau_{{\rm th}}}\left(
%f_\psi-f^{{\rm eq}}_\psi\right),
\end{equation}
where ${\cal C}[f_\psi ]$ is the collision operator of the fermions.
We define $\tau_{{\rm th}}$ is the thermalization time for the fermionic
degrees of freedom, {\it i.e.} the time needed for the fermions of the thermal
bath to response to the change in the background $\phi_c$.
By dividing for small times $0<\delta t < \tau_{{\rm th}}$ the distribution
function into two parts $f_\psi = f^{{\rm eq}}_
{\psi}(\phi_c=0) + \delta f_\psi$ the
lowest order correction $\delta f_\psi$ is determined by the collision operator
with the distributions replaced with $f^{{\rm eq}}_
{\psi}(\phi_c=0)$. The collision
operator can be written in the form
\begin{equation}\label{coll}
{\cal C}[f_\psi ] =
\int {d^3 q \over (2\pi)^3 E_f(q)}{d^3 k \over (2\pi)^3 E_b(k)}
|{\cal M}|^2 f_\psi^{{\rm eq}} ({\bf p})f_\psi^{{\rm eq}} ({\bf q})
f_\phi^{{\rm eq}} ({\bf k})
[e^{\beta \sqrt{ |{\bf k}|^2 + m^2}} - e^{ \beta(|{\bf p }|
+ |{\bf q} |)}].
\end{equation}
The dispersion relations are determined by rapid forward scatterings and thus
the energy conservation relation is now given by
$\sqrt{|{\bf k}|^2 + m^2 + 3\lambda \phi^2} = \sqrt{|{\bf p}|
^2 + g^2\phi^2}
+ \sqrt{|{\bf q}|^2 + g^2\phi^2}$. Applying it, it is straightforward
to convince oneself that the square bracket term in the collision operator
given by (\ref{coll}) gives a contribution of the order $g^2$ or
$\lambda$. Therefore, because also the matrix element is of the order
$g^2$, one obtains the result that
%Now, carefully analyzing Eq. (\ref{bol}) we can infer that for small times
%$0<\delta t<
%\tau_{{\rm th}}$,
\begin{equation}
\frac{f_\psi}
{f^{{\rm eq}}
_\psi(\phi_c=0)}
%\simeq \left[1+\left(\delta t/ \tau_{{\rm th}}\right)\right]=
=1 + {\cal O}\left(g^4\right) +{\cal O}\left(\lambda g^2 \right) .
\end{equation}
On the contrary, expanding the $\phi_c$-dependent equilibrium distribution
\begin{equation}
f^{{\rm eq}}_\psi(\phi_c)=\left(1+{\rm e}^{\beta \sqrt{
\left|{\bf k}\right|^2+g^2\phi_c^2}}
\right)^{-1}
\end{equation}
around $\phi_c=0$, one should get
\begin{equation}
\frac{f^{{\rm eq}}_\psi}{f^{{\rm eq}}
_\psi(\phi_c=0)}= 1+ {\cal O}\left(g^2\right).
\end{equation}
It is therefore clear that for times smaller than
$\tau_{{\rm th}}$, the distribution function $f_\psi$ is far from being
equal to $f^{{\rm eq}}_\psi(\phi_c)$ and, as a consequence, the use of the
equilibrium
one-loop effective potential is {\it not} appropriate to describe the
dynamics of the system.

This example contains the simple, but essential
feature that thermalization takes time and, therefore, the degrees of
freedom of the thermal bath can not always follow the change of the
background $\phi_c$. Indeed, thermalization requires real scattering processes
and therefore is usually fairly slow. Forward scatterings, instead, do
not change the distribution functions of particles traversing
a gas of quanta, but modify their
free dispersion relations. This remains true also in the case of a
nonequilibrium system. Forward scattering manifests itself, for example,
as ensemble and scalar background corrections to the particle masses. Since
the forward scattering rate is usually larger than the non-forward rate,
non-equilibrium ensemble and scalar background corrections are present
even for times smaller than the thermalization time.
Afterwards, non-forward reaction rates become active and the system
thermalizes.

Technically, the thermalization rate $\gamma_{{\rm th}}=\tau_{{\rm th}}^{-1}$
is related to the imaginary part of the two-point function via
$\gamma_{{\rm th}}={\rm Im}\:\Gamma^{(2)}(\omega,{\bf k})/\omega$. Here
$\omega=\omega({\bf k})$ is the solution to the dispersion relations where the
scalar dependent mass is involved.

To decide how fast each particle species do thermalize, one should calculate
the imaginary part of the two point function for each particle species
separately. It is physically, however,  clear that the rates are essentially
the same than their contributions to the Higgs thermalization rate.
Regardless what the actual rates are, the concept of non-equilibrium at
short times remains. In the Standard Model, at
one loop, the imaginary part of the Higgs propagator receives
contributions from decay and absorption (emission) processes. Absorption
(emission) is always proportional to the difference between the
distribution functions of the two external particles in the final and
initial state \cite{iiro}.

At high temperature limit,
%where in the case of light Higgs we have $m^2_H(T)\simeq 0.24\:T^2$,
the thermalization rate of leptons with the Higgs background are
negligible because of their small Yukawa couplings. Also for the quarks the
thermalization rates with the Higgs background are small due to small
Yukawa couplings apart the top quark which has  the rate of the order
$\gamma_t\simeq 10^{-2}\:T$ or smaller.  At two loops
kinematical constraints no longer exist. The leading, purely bosonic,
contribution  has
been estimated in ref. \cite{iiro} from the gauge boson loops and the
thermalization rate for the gauge bosons with the Higgs background
turns out to be at most of order of $\gamma_{{\rm gb}}\simeq
10^{-2}\:T$, whereas the
thermalization rate of the scalar degrees of freedom
receives a further suppression ${\cal O}(\lambda^2/g_2^2)$ with
respect to $\gamma_{{\rm gb}}$,
$\lambda$ being the quartic self-coupling of the Higgs scalar potential
and $g_2$ the $SU(2)_L$ gauge coupling.  Thus, because of both
kinematical and loop suppression factors, the thermalization rate
of fermions, gauge  and scalar bosons with the Higgs background turns
out to be fairly small.

Let us now envisage  the situation in which the $\phi_c$-background,
permeating the thermal bath formed by SM degrees of
freedom, changes from $\phi_c=0$ to a non-vanishing value. From
general point of view the initial field value could be arbitrary with
initial distribution determined by that. In practise the most of the
interesting cases have initially $\phi = 0$ and the applications we consider
use that initial value. A realistic
example for such a situation could be the formation of a subcritical
bubble via thermal fluctuations above the transition temperature $T_f$ or
the passage at a given point of the wall of an expanding critical bubble
nucleated at $T_f$ in a first-order electroweak phase transition.

As said above,
we can not use the imaginary-time formalism to compute the effective
Higgs potential for times smaller than $\sim \gamma^{-1}_{{\rm th}}$
since there is no relation between the density matrix of the system and
the time evolution operator which is of essential importance in the
equilibrium case. There is, however, the real-time formalism of Thermo
Field Dynamics which suites our purposes \cite{tfd} and is characterized
by doubling the degrees of freedom of the heat reservoir.  It is
straightforward to find the (11)-component of the scalar and
fermion propagators for the fields
\begin{eqnarray}
D_{{\rm s}}(k)&=&\frac{i}{k^2-m^2(\phi_c)+i\varepsilon}+2\pi\delta\left[k^2-m^2
(\phi_c)\right]
f_{{\rm b}}(k),\nonumber\\
D_{{\rm f}}(k)&=&\frac{i\left[\not\! k+m(\phi_c)\right]}
{k^2-m^2(\phi_c)+i\varepsilon}-2\pi\delta\left[k^2-m^2(\phi_c)\right]
\left[\not\! k+m(\phi_c)\right]
f_{{\rm f}}(k),\nonumber\\
\end{eqnarray}
and analogous formulae for the gauge boson propagators. Note, that here
$f_{{\rm b,f}}(k)$ are arbitrary distribution functions restricted by
requiring  the number expectation value to be positive.

The general expression for the thermal part of the effective potential
can be calculated by usual way (e.g. by the tadpole method)
but the calculation leads now to a different expression for the potential
than Eq. (\ref{pot}) (with $f^i_{{\rm eq}}$ replaced by the generic
expression $f_{{\rm b}({\rm f})}$). The derivation of formula (\ref{pot})
uses explicitly the concept of thermal equilibrium \cite{dolan} or,
as in the case of tadpole method in real time formalism, it is assumed
that the distribution is a function of energy, i.e. it is $\phi_c$ dependent.
Now, for times smaller than
$\gamma^{-1}_{{\rm th}}$, $f_{{\rm b},{\rm f}}$ are well approximated
by
\begin{equation}
\label{ansatz}
f^{{\rm eq}}_{{\rm b},{\rm f}}=\left({\rm e}^{\sqrt{|{\bf k}|^2+
m^2(T,\phi_c=0)}/T}\mp 1\right)
^{-1},
\end{equation}
where  $m(T,\phi_c)$ is
the plasma mass at finite temperature for each degree of freedom.

The choice given in Eq. (\ref{ansatz}) for the distribution functions
needs some justification: at $t<\gamma^{-1}_{{\rm
th}}$ particles have not had time enough to have real scatterings, and
thus the distributions have not yet had time enough to feel the change of the
background $\phi_c$. So they remains the same as they were when the
background mass was
$\phi_c=0$. Nevertheless, since forward scattering are much faster than
non-forward reaction rates, the free dispersion relation of particles
get modified  and their masses receive plasma corrections from the
ensemble. Furthermore, since quanta interactions are rather fast, {\it
e.g.} mediated by strong force, the use of the equilibrium distribution
function is well motivated. We refer the reader to ref. \cite{second}
for more details.

%Plugging the distribution functions (\ref{ansatz}) into the expression
%(\ref{pot}) for
Calculating the effective potential, we find
\begin{equation}
V\left(T,\phi_c\right)=\frac{1}{2\pi^2}%\sum_i\:
{\rm Tr} \: \int_0^\infty\:d|{\bf k}|\:|{\bf k}|^2\:
\sqrt{|{\bf k}|^2+m^2(T,\phi_c)}\:
\left({\rm e}^{\sqrt{|{\bf k}|^2+m^2(T,\phi_c=0)}/T}\mp 1\right)^{-1},
\end{equation}
where the trace is taken over all degrees of freedom and $m^2(T, \phi_c)$ is
a general mass matrix. Same techniques as used in the calculation of
equilibrium potential \cite{dolan}
can be applied here resulting a general formula
in which the leading terms are included
\begin{equation}\label{general}
V\left(T,\phi_c\right) = {T^4\over 2\pi^2} \left[\:
{\rm Tr}\: I_{{\rm b}}\left({m_{{\rm b}}(T,\phi_c)\over T}, {m_{{\rm
b}}(T,\phi_c=0)\over T}\right) +
{\rm Tr}\: I_{{\rm f}}\left({m_{{\rm f}}(T,\phi_c)\over T},
{m_{{\rm f}}(T,\phi_c=0)\over T}\right)\right].
\end{equation}
The bosonic (fermionic) functions $I_{{\rm b}({\rm f})}$
are given by
\begin{equation}
I_{\rm b}(x, y) =
\frac 12 \left[\frac 16 \pi^2 - y + \frac 14 y^2\right]\:x^2 - \frac 13 x^3 +
\frac 1{16} \left[ \frac 34 - 1 - \gamma_E + \ln (4\pi )\right]\:x^4
- \frac 1{32} x^4
\ln x^2 + \dots
\end{equation}
and
\begin{equation}
I_{\rm f}(x, y) =
\left[\frac 1{24} \pi^2 - \frac 18 y^2\right]\:x^2 +
\frac 1{16} \left[ \frac 14 + \gamma_E - \ln (\pi )\right]\:x^4
+ \frac 1{32} x^4\ln x^2
+ \dots
\end{equation}
Here $\gamma_E$ is the Euler constant and $m_{{\rm b}({\rm f})}$
is a general mass matrix
of bosonic (fermionic) degrees of freedom. In the formulas above the dots
stand for higher powers of $x$. It is worth to note that in the mass matrices
we are able to use the temperature dependent masses: the resummation
can be done similarly than in the equilibrium case \cite{quiros}.
This potential clearly
differs from the usual equilibrium potential, whereas the zero-temperature part
remains naturally unaltered. In particular, the next-to-leading
boson contribution coming from the infrared region to the cubic term
differs from the equilibrum one by a factor $2/\pi$.

After adding up the two parts
we have plotted the effective potential for the Standard Model
in the particular case of $m_H=65$ GeV, Fig. 1. The critical temperature
$T_c$ defined as the temperature
at which $V(T_c,0)=V(T_c,\phi_c^+)$, $\phi_c^+$ being the
non-vanishing minimum, is around 94.27 GeV, higher than the critical
temperature for the equilibrium case (around 90 GeV) for the same choice
of the Higgs mass, see \cite{second} for a detail analysis.

Let us now briefly discuss two physical cases in which our
nonequilibrium approach to the  effective potential
might qualitatively effect the
conclusions commonly drawn employing the equilibrium effective
potential.

As already mentioned, the role played by subcritical bubbles at the onset of
a first-order electroweak phase transition is still disputed \cite{sb}.
It might be possible that the amplitude of thermal fluctuations be so
large that the fraction occupied by the symmetric minimum in the
neighborhood of the critical temperature $T_c$ becomes of order unity,
thus preventing the formation of expanding critical bubbles and
hindering any mechanism for electroweak baryogenesis.

To compute the average amplitude of thermal fluctuations it is commonly
hypothesized that the free energy ${\rm F}$ of a subcritical bubble
configuration
$\phi_{{\rm sb}}$ receives  contribution
from the {\it equilibrium} effective
potential. This  gives rise to a Boltzmann weight $\sim
{\rm exp}\left[-{\rm F}(\phi_{{\rm sb}})/T\right]$
for each configuration.  However, subcritical
bubbles, being unstable objects, tend to shrink. It is rather
conceivable that thermal fluctuations are dominated by large-size
configurations whose lifetime can be roughly estimated to be $\tau_{{\rm
sh}}\sim m^{-1}_H(T)$. Since $\tau_{{\rm
sh}}\ll \gamma_{{\rm th}}^{-1}$ at the critical temperature, particles
inside the subcritical bubble do not experience thermalization with the
$\phi_{{\rm sb}}$-configuration. Consequently, the equilibrium effective
potential is completely inadequate to determine the free energy of such
configurations and must be taken over by the nonequilibrium one.

Our nonequilibrium approach might also have consequences for the
determination of the velocity $v_{{\rm w}}$ and width $L_{{\rm w}}$ of
critical bubble walls expanding in the thermal bath at the onset of a
first-order electroweak phase transition.
The determination of these parameters has  received much attention
\cite{v} since the discovery of the possibility to generate the baryon
asymmetry at the electroweak scale. Intuitively speaking, the velocity
and the shape of the bubble wall depend upon two factors: the pressure
difference $\Delta p$ at the two edges of the bubble wall ({\it i.e.}
between the broken and the unbroken phase), which allows for the
expansion, and the friction force due to the collisions of the plasma
particles off the wall. In the previous treatments, the population density
$f^i$ of each species $i$ has been splitted in the {\it equilibrium} one
plus a small deviation, $f^i=f^i_{{\rm eq}}(\phi_c)+\delta f^i$. The
pressure difference $\Delta p$ is then determined by $\Delta V_{{\rm
eq}}(\phi_c)$ and the friction force arises due to departure from
the thermal equilibrium distribution. However, since thermalization
between the degrees of freedom of the thermal bath
and the bubble wall background takes a finite time, this picture is
correct only if $\gamma^{-1}_{{\rm th}}\ll (L_{{\rm w}}/v_{{\rm w}})$.
If the opposite limit, equilibrium is not attained in
the vicinity of the bubble wall in the broken phase and the
nonequilibrium effective potential should play a role in estimating the
pressure difference $\Delta p$ between the two phases.

We are currently investigating both issues \cite{second}.

\acknowledgments

The authors thank the organizers of the TAN Workshop held in Vigs{\o}e,
August 1995, where part of this work was done.

\newpage
{\large\bf Figure Captions}
\vspace{1 cm}

Fig. 1: The Standard Model nonequilibrium effective potential (in units
of GeV)
for the particular choice $m_H=65$ GeV.

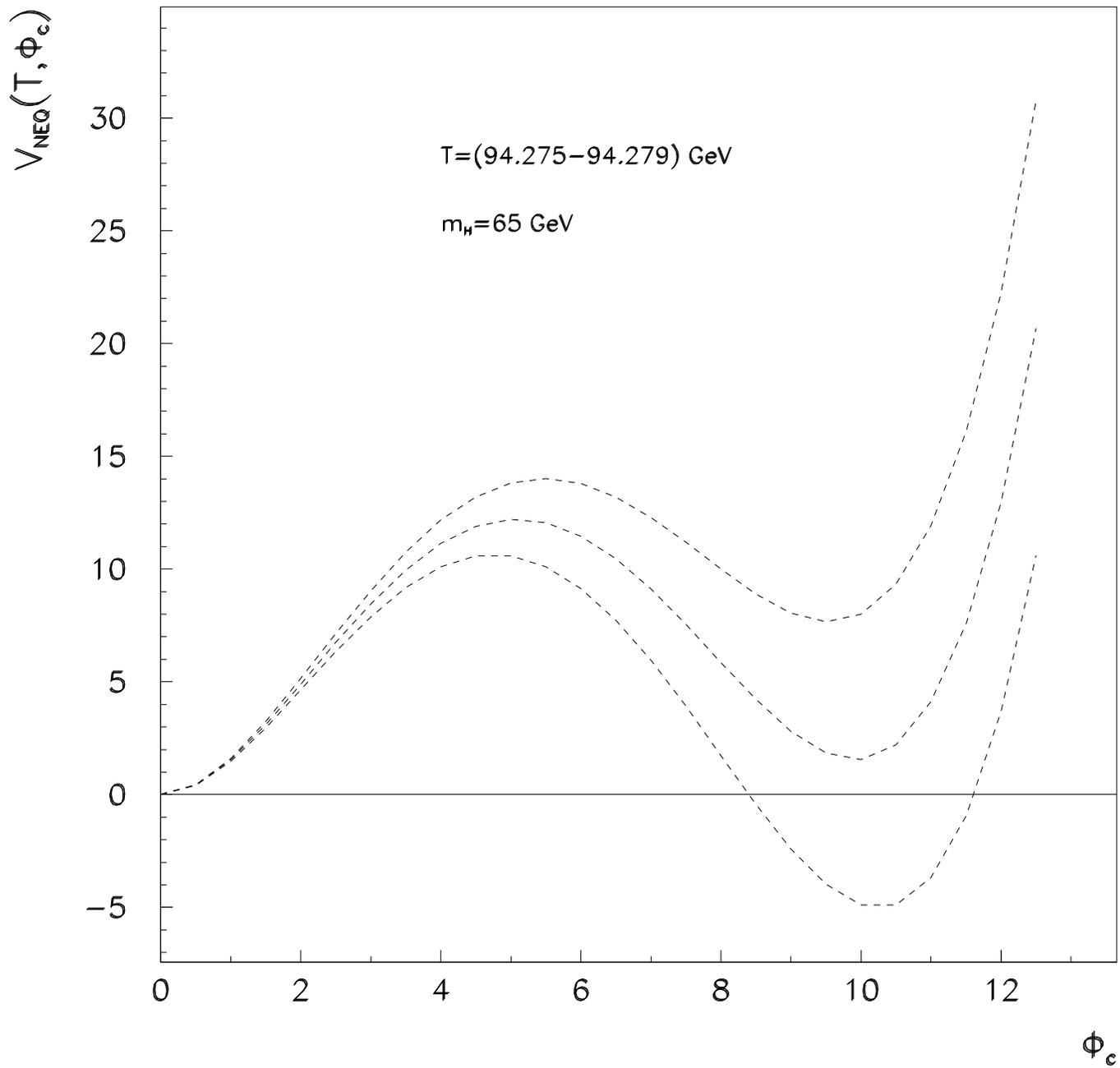
\end{document}